\documentclass[twocolumn,conference]{IEEEtran}
\usepackage[T1]{fontenc}
\usepackage[latin9]{inputenc}
\usepackage{color}
\usepackage{enumitem}
\usepackage{amsmath}
\usepackage{amssymb}
\usepackage{graphicx}
\usepackage{wasysym}
\usepackage{esint}
\usepackage[unicode=true,
 bookmarks=true,bookmarksnumbered=true,bookmarksopen=true,bookmarksopenlevel=1,
 breaklinks=false,pdfborder={0 0 0},pdfborderstyle={},backref=false,colorlinks=false]
 {hyperref}
\hypersetup{pdftitle={Your Title},
 pdfauthor={Your Name},
 pdfpagelayout=OneColumn, pdfnewwindow=true, pdfstartview=XYZ, plainpages=false}

\makeatletter

\providecommand{\tabularnewline}{\\}

\usepackage{graphicx}
\usepackage[caption=false,font=footnotesize]{subfig}
\usepackage{algorithm,algpseudocode}
\usepackage{remreset}
\usepackage[noadjust]{cite}

\@ifundefined{showcaptionsetup}{}{%
 \PassOptionsToPackage{caption=false}{subfig}}
\usepackage{subfig}
\makeatother

\begin{document}
\title{Latency of Concatenating Unlicensed LPWAN with Cellular IoT: An Experimental
QoE Study}
\author{Alvin~Ramoutar, Zohreh~Motamedi and Mouhamed~Abdulla\\
School of Electrical Engineering, Faculty of Applied Science and Tech.\\
Sheridan Institute of Technology, Toronto, Ontario, Canada\\
Email: \{ramoutal,\,motamedz,\,abdulmou\}@sheridanc.on.ca}
\maketitle
\begin{abstract}
Developing low-power wide-area network (LPWAN) solutions that are
efficient to adopt, deploy and maintain are vital for smart cities.
The poor quality-of-service of unlicensed LPWAN, and the high service
cost of LTE-M/NB-IoT are key disadvantages of these technologies.
Concatenating unlicensed with licensed LPWANs can overcome these limitations
and harness their benefits. However, a concatenated LPWAN architecture
will inevitably result in excess latency which may impact users' quality-of-experience
(QoE). To evaluate the real-life feasibility of this system, we first
propose a concatenated LPWAN architecture and experimentally measure
the statistics of end-to-end (E2E) latencies. The concatenated delay
margin is determined by benchmarking the latencies with different
LPWAN architecture schemes, namely with unlicensed IoT (standalone
LoRa), cellular IoT (standalone LTE-M), and concatenated IoT (LoRa
interfaced with LTE-M). Through extensive experimental measurement
campaigns of $30,000$ data points of E2E latencies, we show that
the excess delay due to LPWAN interfacing introduces on average less
than $300$ milliseconds. The proof-of-concept results suggest that
the latency for concatenating unlicensed LPWAN with cellular IoT is
negligible for smart city use cases where human perception and decision
making is in the loop.
\end{abstract}


\IEEEpeerreviewmaketitle{}

\section{Introduction}

IoT applications for enhanced and massive machine type communications
(eMTC/mMTC) is growing at an unprecedented rate, empowering connectivity
in various fields such as healthcare, agriculture, climate and smart
city applications. To support communications for various IoT use cases,
low-power wide-area network (LPWAN) technologies are used where appropriate
to transmit small payloads at long distances with minimal power consumption.

\begin{figure}[t]
\begin{centering}
\includegraphics[width=0.97\columnwidth]{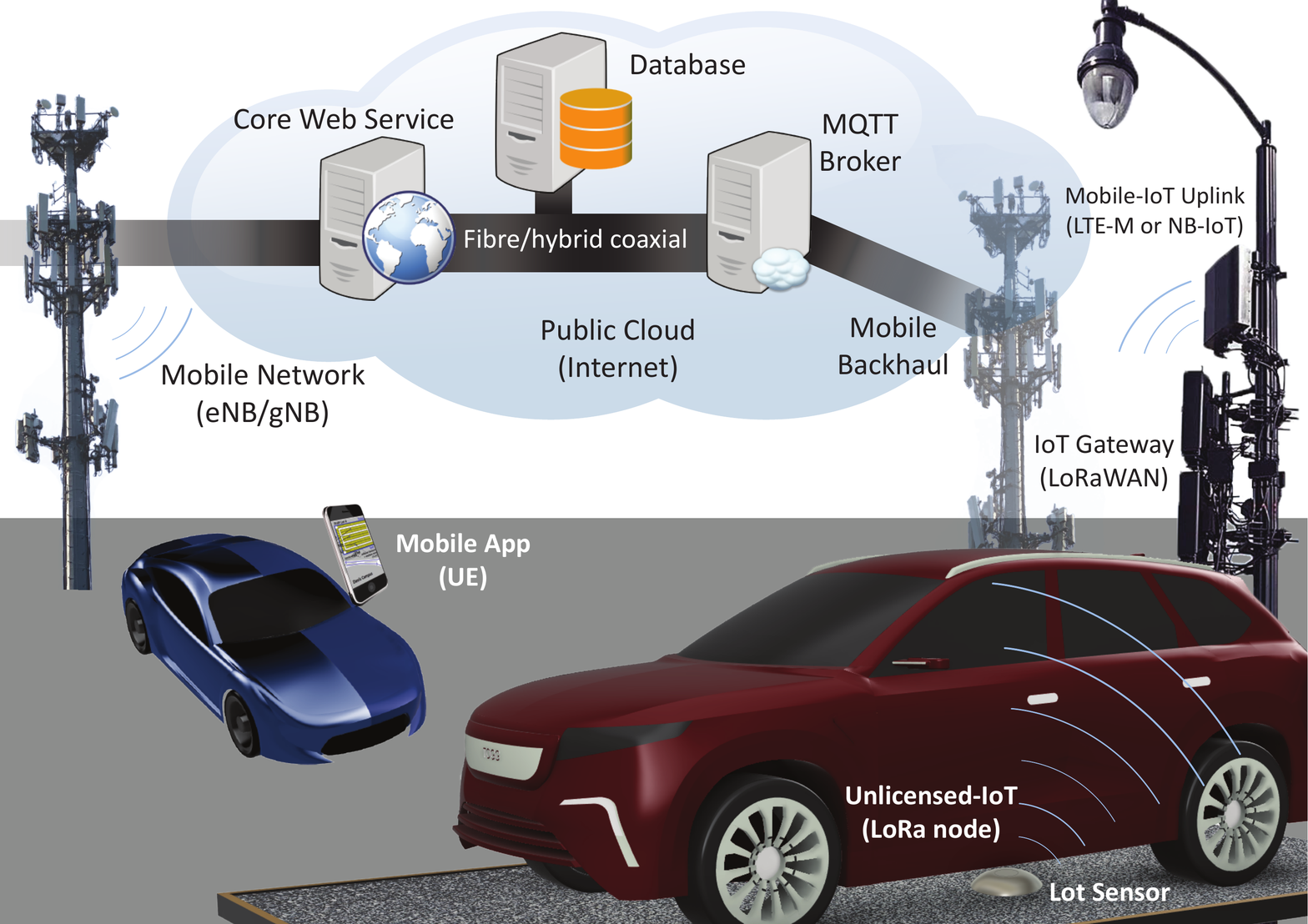}
\par\end{centering}
\vspace{-0.2cm}

\begin{centering}
\caption{Application of system concatenation for smart cities with intelligent
parking using unlicensed LPWAN backhauled via cellular IoT uplink.
\label{Fig1. Smart_Parking}}
\par\end{centering}
\vspace{-0.5cm}
\end{figure}

Traditional LPWAN, such as LoRa/LoRaWAN and Sigfox, operate in unlicensed
radio spectrum. While freely available on ISM bands, unlicensed IoT
is prone to RF interference and poor quality-of-service (QoS). On
the other hand, licensed IoT offers a more reliable communication
as it depends on cellular infrastructure deployed with careful network
planning. As a consequence, the industry is complementing the IoT
ecosystem with licensed LPWAN for eMTC and mMTC applications. Today,
this is done through the use of LTE-M and NB-IoT technologies defined
by 3GPP\,rel.\,13-14 (4G), and rel.\,15-16+ (5G). Granted, licensed
IoT will require regular subscription for each sensor node with a
mobile network operator, and this is a costly solution for large-scale
smart city deployment.

To overcome these limitations, we propose a concatenated network architecture
that harnesses the advantages of both unlicensed and licensed LPWAN.
Consider the smart parking application shown in Fig.~\ref{Fig1. Smart_Parking},
where large or massively deployed unlicensed IoT sensors detect status
changes in parking spaces and backhauls aggregated data via a number
of cellular IoT nodes to the cloud for user access. Although this
concatenated LPWAN architecture combines affordability and reliability
together, the main drawback is the added latency due to interfacing
and data hopping of information packets.

In literature, various contributions have studied the throughput,
range, and power of unlicensed (e.g., \cite{magrin.d+:2017:loraperf,rochester.e.m+:2020:loraperf,furtado.a+:2020:loraperf})
and licensed IoT (e.g., \cite{elfawal.a.h+:2018:ltemperf,azari.a+:2020:nbiotperf,sulyman.a.i+:2019:ltemnbiotperf}).
Furthermore, a monitoring system was tested by combining NB-IoT and
LoRa \cite{zhang.x+:2019:monitoring}, and a hybrid 3G-LoRa-Sigfox
for power grid monitoring was verified with legacy technologies \cite{campo.g.d+:2019:monitoring}.
While these works provide valuable insights, a proof-of-concept (PoC)
and quality of experience (QoE) assessment with end-to-end (E2E) latency
measurement for concatenated LPWAN architecture is yet to be explored.

In this paper, we propose a concatenated network architecture and
protocols with unlicensed LPWAN and cellular IoT over a private core
infrastructure. Due to the availability of 3GPP LTE Cat M1 service
by local mobile operators, we implement the architecture with LoRa
and LTE-M networks. Through E2E latency measurement campaigns, $30,000$
data points are experimentally collected and statistically analyzed
for standalone and concatenated LPWAN architectures \cite{Dataset2021}.
Data-driven discussions and remarks on the feasibility of a concatenated
LPWAN architecture to achieve a target QoE for smart city applications
with users in the loop is also provided.

\section{\label{Sec2 - Network Architecture}Network Architecture}

The E2E architecture from in-field sensing to remote detection is
based on three major steps: (i) uplink communications, (ii) cloud
and core infrastructure, and (iii) mobile end-user application. Our
proposed concatenated LPWAN architecture with different possible IoT
technology choices is shown in Fig.~\ref{Fig2. Concatenated_LPWAN_Architecture}.
As seen first, information packets from large or massive amount of
sensor nodes are aggregated by fewer IoT gateways. Then, the aggregated
data from gateways is relayed to cellular IoT nodes for uplink communication
to a mobile carrier network. The data from packets is then processed,
stored, and made available to mobile clients from the core infrastructure
residing in either a public or private cloud. An explanation of the
various sub-components of this concatenated architecture is provided
below.

\subsection{Uplink Communications}

\noindent Starting with IoT sensor nodes, this subsection covers communications
up to the public cloud.

\makeatletter
\@removefromreset{subsubsection}{subsection}
\renewcommand\thesubsubsection{\arabic{subsubsection}}
\makeatother

\vspace{0.05cm}

\subsubsection{Data aggregation of IoT sensors}

Payload construction by a sensor IoT node begins after new sensing
data is detected and produced. The payload is then split into byte
segments and transmitted over radio on unlicensed spectrum via LoRa
systems. Developed by Semtech, LoRa is a physical (PHY) and data link
(MAC) layers protocol utilizing chirp spread spectrum modulation to
transmit between $0.3$ and $50$\,kbps over large distances with
minimal power consumption. In North America, LoRa operates on the
unlicensed $902$ to $928$\,MHz ISM frequency band. Often paired
with a LoRa system is LoRaWAN, a network  layer protocol used to
route data from LoRa nodes to a gateway and then to the internet.
From end-user device to network server and at the application level,
LoRaWAN uses two layers of 128 bits AES encryption \cite{LoRa_Standard}.

\vspace{0.05cm}

\subsubsection{Gateway to cellular IoT communication}

Once the transmitted sensor payloads are received by an unlicensed
gateway, they are first extracted from the LoRa protocol and then
sent via a serial protocol to a licensed cellular IoT radio within
the same device package. At this same point, the payload is wrapped
in a lightweight messaging protocol such as MQTT in preparation for
IP routing.

\vspace{0.05cm}

\subsubsection{Transmission from cellular IoT to mobile carrier}

\begin{figure*}[t]
\begin{centering}
\includegraphics[width=0.85\paperwidth]{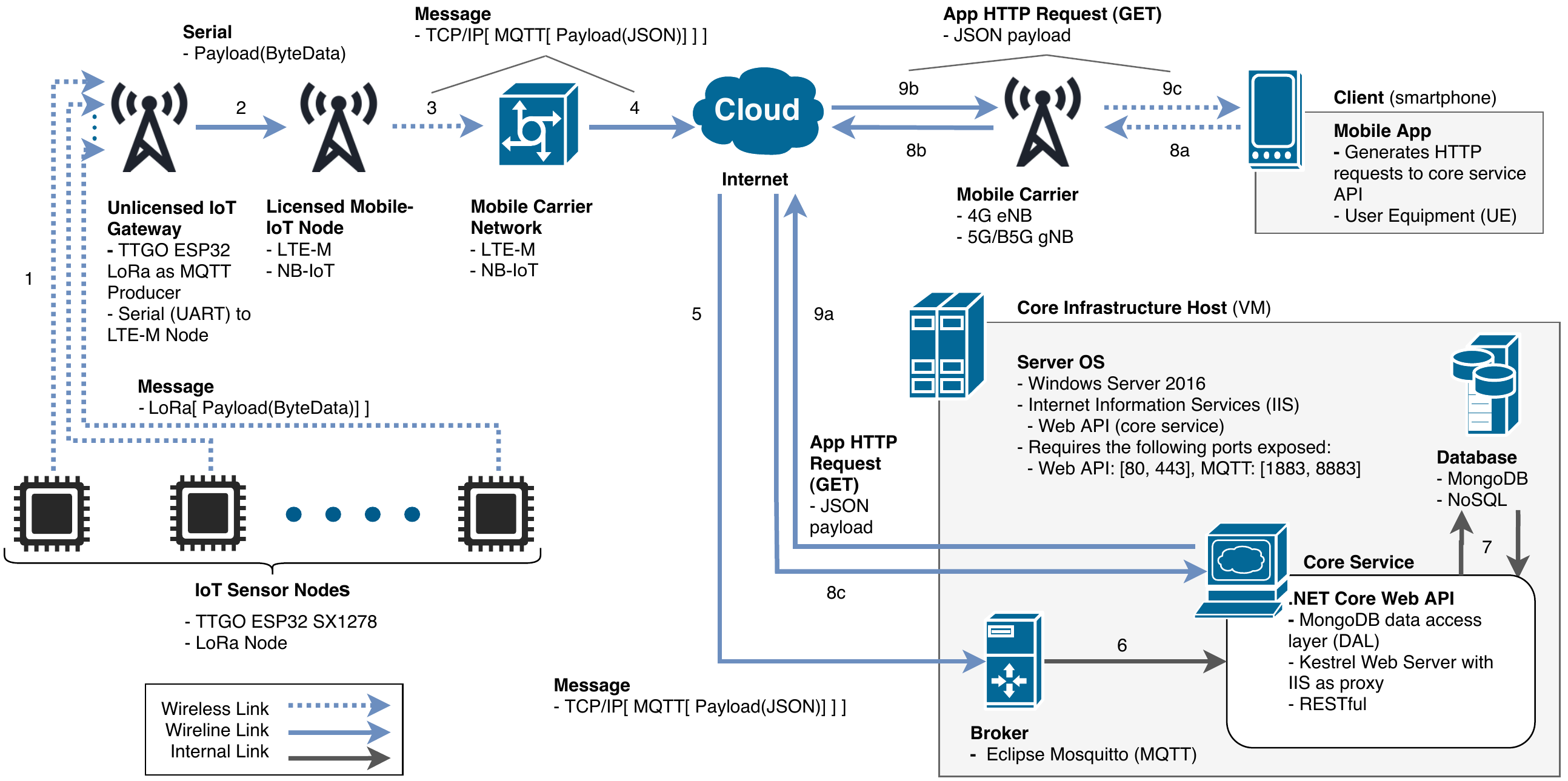}
\par\end{centering}
\vspace{-0.1cm}

\caption{Proposed E2E concatenated LPWAN network architecture with unlicensed
IoT (LoRa), cellular IoT (LTE-M or NB-IoT) and a private core infrastructure.
\label{Fig2. Concatenated_LPWAN_Architecture}}
\end{figure*}

This communication serves as the backhaul, forwarding payload via
licensed cellular IoT. LTE Cat M1 or NB-IoT devices will then uplink
to nearby eNB or gNB cell towers to connect to the rest of the mobile
carrier network. The concatenated architecture of Fig.~\ref{Fig2. Concatenated_LPWAN_Architecture}
is compatible and it can be adapted with both LTE-M and NB-IoT. In
reality, the regional availability of cellular IoT by mobile operators
will dictate which of these licensed LPWAN options is accessible by
users. In general, NB-IoT service is widely available in Asia and
Europe. Currently in North America, LTE-M service is more common,
and NB-IoT is still in its early stages. Developed by 3GPP rel. 13+
for IoT applications, LTE-M operates on closed network servers and
on several licensed frequencies (e.g., bands 2, 4, 5, 12 and 13 in
Canada). This technology uses time division duplex and a peak data
rate of 1 Mbps for uplink and downlink with $1.4$ MHz bandwidth.
Since each cellular IoT device requires a SIM card and service contract,
large-scale implementation of LTE-M is expensive. The proposed concatenated
architecture alleviates the need for a costly deployment of large
or massive volume of cellular IoT nodes through system interfacing
with unlicensed LPWAN sensors.

\vspace{0.05cm}

\subsubsection{Communication from evolved packet core to cloud}

The IP uplink is made possible by the service provider which routes
data from the mobile carrier network through wireline comprising of
hybrid coaxial fiber. Eventually, data packets reaching the evolved
packet core are forwarded to the cloud for computation and transmission.

\vspace{-1bp}

\vspace{-1bp}

\subsection{Cloud and Core Infrastructure}

\noindent IoT sensor payloads are transferred through the internet
to the core infrastructure for processing.

\vspace{0.05cm}

\subsubsection{TCP/IP routing to core infrastructure}

The communication involves routing from the carrier network through
IP communication to a broker hosted by a cloud infrastructure. This
architecture identifies IoT sensor nodes and gateway technology as
part of the private cloud receiving MQTT payloads from concentrated
LoRa traffic. A private core infrastructure could be substituted
by a public cloud (e.g., Azure or AWS IoT). Rather than involved elaborate
setups, public cloud solutions facilitate insights into the analytics
of sensor payloads. For instance, Azure Monitor and AWS Cloud Watch
both enable resource metric and telemetry collection while visualizations
are generated by Microsoft Power BI and AWS Quicksight. Although developers
may invest effort into implementing these services for robust security,
cloud native IoT platforms deliver them automatically as pay-per-use
subscriptions which often scale quickly \cite{pierleoni.p+:2020:cloudiot}.
The cost and offering implications make public clouds viable for corporate
IoT roll-out with private clouds more suitable and secure for testing
and research.

\vspace{0.05cm}

\subsubsection{Delivering data from message broker to core service}

To address high traffic from large-scale IoT sensors, MQTT messages
are ingress by a broker such as Eclipse Mosquitto. Similar to publisher/subscriber
paradigms, the broker delivers sensor payload to the core service
via a subscribed channel/topic. The core service is an application
that serves three main objectives: business logic processing, data
access layer (DAL), and application programming interface (API). They
all can be decoupled for a service-oriented architecture.

\vspace{0.05cm}

\subsubsection{Storing sensor information in core service database}

The nature of IoT oftentimes results in unstructured data, which is
best suited for NoSQL data stores that can leverage a variety of data
models from tables to documents. The key benefit of going with document-based
database models would be alignment with common and modern web formats
and DALs. One such technology that achieves this is MongoDB, where
documents are stored in a superset of JSON known as BSON.

\vspace{-1bp}

\vspace{-1bp}

\subsection{Mobile Application and Core Service}

\noindent The consumption of IoT sensor data by clients is identified
in this final subsection of the architecture.

\vspace{0.05cm}

\subsubsection{Mobile client request}

In this concatenated LPWAN network architecture, the user equipment
(UE) performs an action requesting data over Hypertext Transfer Protocol
(HTTP). This request is sent through the UE\textquoteright s connected
network, eventually reaching the internet. Once it reaches the core
services, the request is routed to the core service API.

\vspace{0.05cm}

\subsubsection{Core service response}

The core service follows the representational state transfer (REST)
architecture for web service communications. RESTful APIs are ideal
over simple object access protocol (SOAP) and others since it not
only provides clear separation between data producers and consumers,
but it is also data format agnostic. This enables developers to create
highly compatible APIs to consume and deliver a variety of IoT sensor
data with minimal coupling to UE's application. While REST is modern,
it is unsuitable and often unnecessary for certain applications. For
instance, when a core service must constantly stream data to a client
or vice-versa, a persistent bi-directional communication channel should
be established. WebSocket is one protocol that serves this purpose,
creating a TCP duplex connection that enables message pooling without
the need to wait for a response like HTTP \cite{vujovic.m+:2015:websocket}.

\section{\label{Sec3 - Experimental Setup}Experimental Setup and Methodology}

\begin{figure*}[t]
\begin{centering}
\includegraphics[width=0.85\paperwidth]{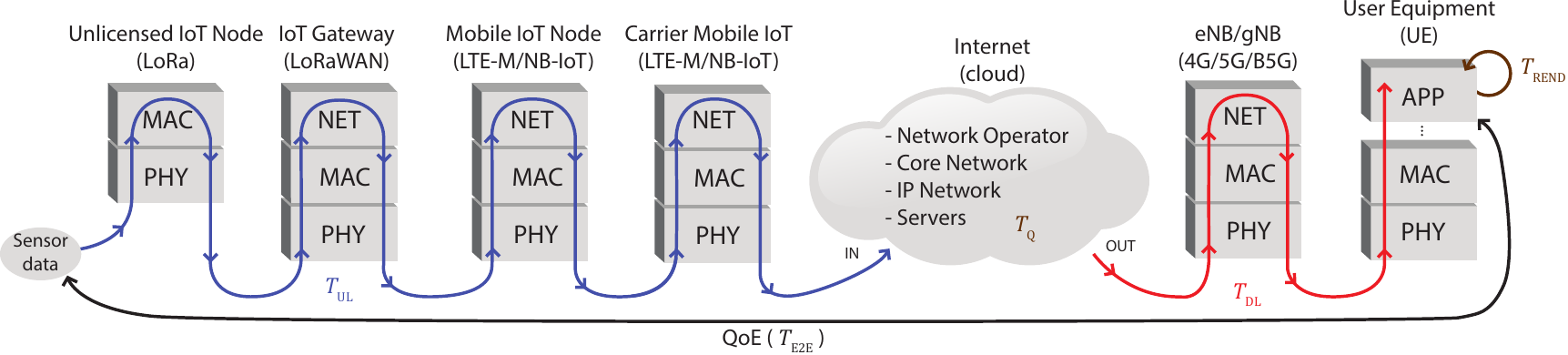}
\par\end{centering}
\vspace{-0.2cm}

\caption{Trace of E2E latency measurement of data from sensing to UE's application
passing across the different OSI layers of the network architecture.
 \label{Fig3. OSI_Latency}}
\end{figure*}

As a critical key performance indicator (KPI) for QoE assessment,
we experimentally evaluate a precise E2E latency for the concatenated
LPWAN architecture with LoRa and LTE-M networks. Using the architecture
detailed in Sec.\,\ref{Sec2 - Network Architecture}, we divide the
E2E latency $T_{{\rm E2E}}$ into several components based on where
we have measurable insight with data entry and exit of architecture
subsystems to capture timing information. These components are defined
as follows:
\begin{itemize}
\item $T_{{\rm UL}}$: uplink communication latency of IoT sensor payload
to the core infrastructure (Fig.~\ref{Fig2. Concatenated_LPWAN_Architecture},
links\,1\,to\,7).
\item $T_{{\rm Q}}$: queuing delay in the core service database before
client request (Fig.~\ref{Fig2. Concatenated_LPWAN_Architecture},
core infrastructure).
\item $T_{{\rm DL}}$: mobile client request and resulting core service
downlink response latency (Fig.~\ref{Fig2. Concatenated_LPWAN_Architecture},
links\,8\,and\,9).
\item $T_{{\rm REND}}$: internal latency for graphical user interface (GUI)
rendering at the UE (Fig.~\ref{Fig2. Concatenated_LPWAN_Architecture},
client's application).
\end{itemize}
For further clarification, the time measures are also shown in Fig.~\ref{Fig3. OSI_Latency}
as the sensing packets pass across different OSI layers and network
components of the architecture. The resultant E2E latency is represented
as the sum of the time measures; namely
\begin{align}
 & T_{{\rm E2E}}=T_{{\rm UL}}+T_{{\rm Q}}+T_{{\rm DL}}+T_{{\rm REND}}\text{ .}\label{eq1: Latency}
\end{align}

\begin{figure}[tbh]
\begin{centering}
\includegraphics[width=1\columnwidth]{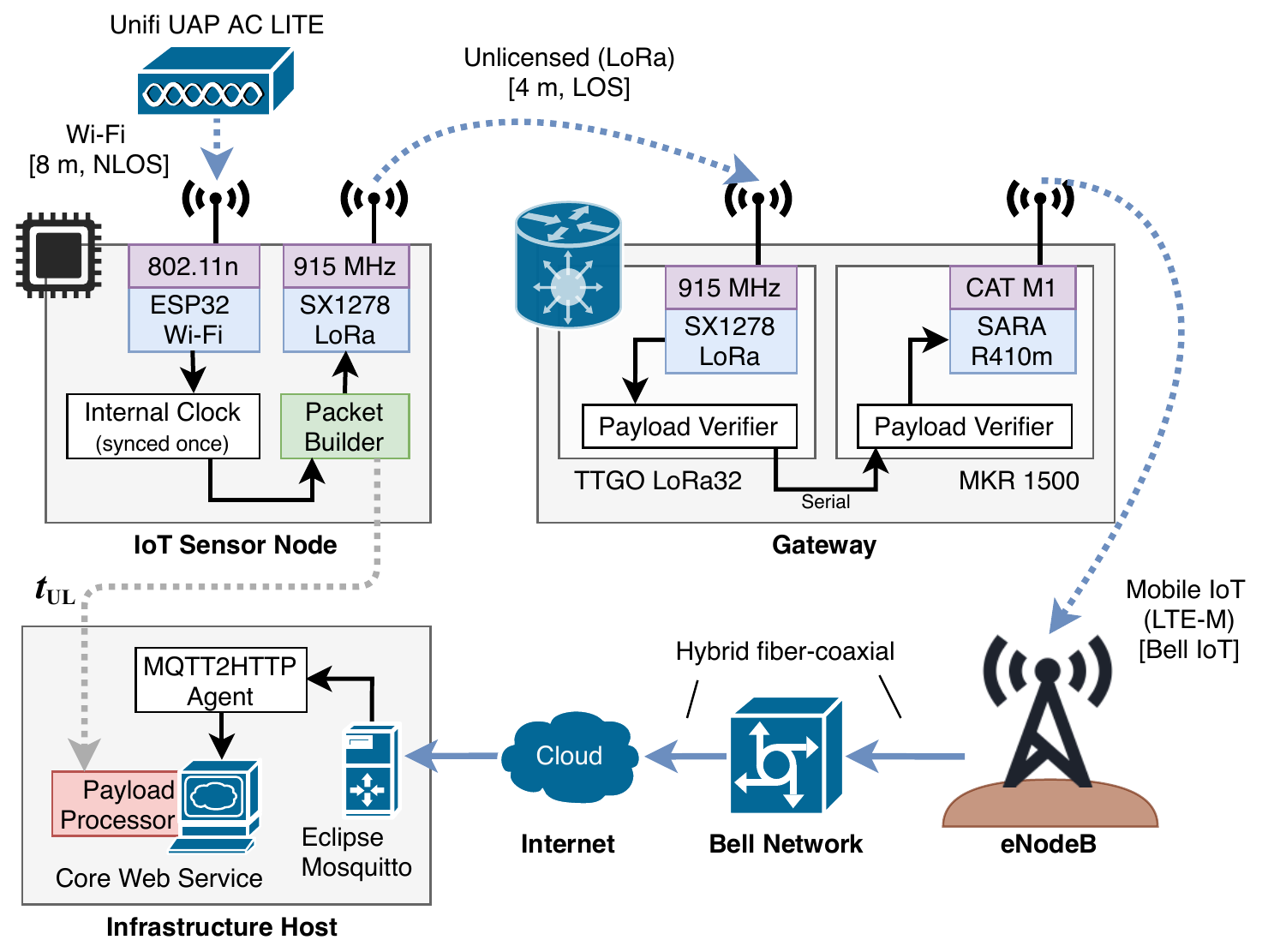}
\par\end{centering}
\vspace{-0.2cm}

\caption{Experimental setup for data collection of E2E latency for the concatenated
LPWAN network architecture. \label{Fig4. Experimental_Setup}}

\vspace{-0.5cm}
\end{figure}

\begin{figure}[t]
\begin{centering}
\includegraphics[width=0.92\columnwidth]{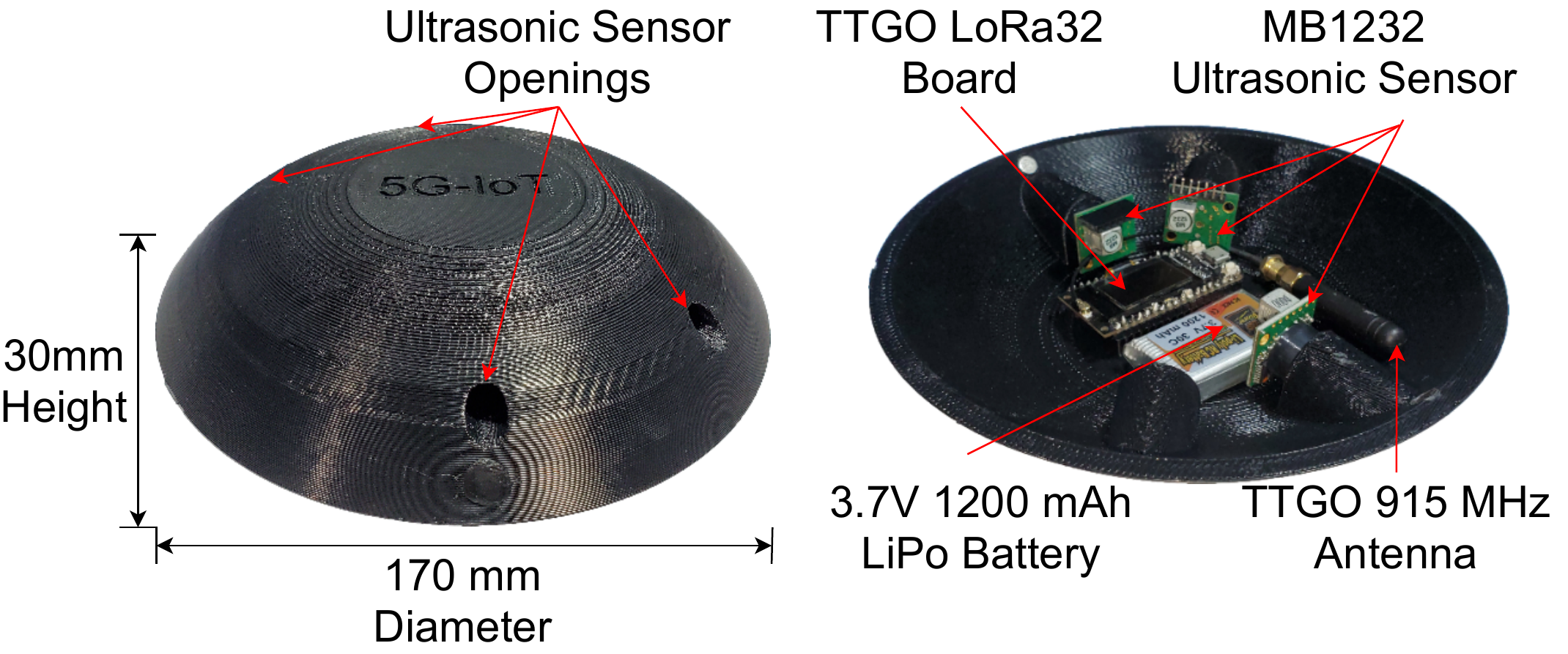}
\par\end{centering}
\vspace{-0.3cm}

\caption{Smart parking LPWAN sensor with four MaxBotix ultrasonic narrow beam
sensor module with up to 40\,Hz read rate, and a TTGO LoRa32 microcontroller
with a 2\,dBi omnidirectional antenna operating at 915\,MHz. \label{Fig5. Smart_Parking_Sensor_Casing}}

\vspace{-0.5cm}
\end{figure}

The experimental setup that implements the proposed concatenated LPWAN
architecture is detailed in Fig.~\ref{Fig4. Experimental_Setup},
where the key specifications for the communication system and network
equipment considered for the data collection campaign are listed in
Table~\ref{ Tab I. HW specifications}. The PoC experiment was conducted
in an outdoor parking lot in the suburbs of the Greater Toronto Area
using the smart parking LPWAN sensor node shown in Fig.~\ref{Fig5. Smart_Parking_Sensor_Casing}.
This particular sensor is intended for efficient deployment and use
at the center of a 2-by-2 parking lot configuration. Contained within
this integrated unit is a TTGO LoRa32 microcontroller that processes
data from four MB1232 ultrasonic sensors. We designed the sensor with
openings on the top-side of the casing where the angles of the port
holes was determined based on North American parking lot dimensions
with each ultrasonic sensor pointing towards the center of a parking
spot.

During initialization, the LPWAN sensor node with TTGO ESP32 LoRa
development board establishes a $2.4$\,GHz radio link connectivity
once over IEEE 802.11n. Time is synchronized via network time protocol
(NTP) server before disconnecting. LoRa packets of $28$\,bytes are
then built with a $2$\,bytes label, 10\,bytes Unix timestamp, 15\,bytes
static padding for sensor data fields, including device identification
and token-based authentication (auth token), and finally with a null
terminator byte to indicate the end of the packet. Timing for $T_{{\rm UL}}$
begins immediately once a new data packet is transmitted by a LoRa
sensor node. To gather a large sample size of latency measurements,
the process is repeated every 500\,ms, with data flowing at a rate
of $28\times8/500\text{ ms}=0.448\text{ kps}$.

Another TTGO ESP32 development board is used as a LoRa gateway that
scans for a payload of 28\,bytes and validates its integrity against
a 2\,bytes label. Once validated, the payload is sent serially to
the MKR\,1500 which assembles an MQTT UDP packet. The MKR transmits
the packet via LTE-M through Bell Canada's network to its final destination
at the core infrastructure host that runs the Eclipse Mosquitto broker.
Subscribed to this channel/topic is a custom lightweight service agent
MQTT2HTTP forwarding to the core service API. Measurement of $T_{{\rm UL}}$
concludes when the payload is received at the core service, its timestamp
difference calculated, and a persistent entry written to MongoDB database.

\begin{table}[b]
\vspace{-0.2cm}

\caption{Specifications of  communication system and network components used
for empirical measurements \label{ Tab I. HW specifications}}

\vspace{-0.2cm}

\resizebox{0.5\textwidth}{!}{%
\centering{}{\scriptsize{}}%
\begin{tabular}{c||ccc}
\hline 
\textbf{hardware equipment} & \textbf{CPU/MPU} & \textbf{storage} & \textbf{radio access}\tabularnewline
\hline 
\hline 
TTGO LoRa32 & ESP32 & $520$\,kB & $915$\,MHz, $150$\,Mbps\tabularnewline
(LoRa/LoRaWAN) & DOWDQ6 & SRAM & $20$\,dBm (Tx), $-148$\,dBm (Rx)\tabularnewline
\hline 
MKR NB 1500 & SAMD21 & $256$\,kB & Bands\,4,\,5,\,12, $375$\,kbps (UL)\tabularnewline
(LTE Cat M1) & Cortex & SRAM & $23$\,dBm (Tx), $-105$\,dBm (Rx)\tabularnewline
\hline 
HPE ProLiant DL380 G5 & Intel Xeon & $40$\,GB disk & Broadcom NC373i\tabularnewline
(core infrastructure) & E5440 & $4$\,GB RAM & Gigabit server adapter\tabularnewline
\hline 
Samsung Galaxy S7 & Exynos 8890 & $32$\,GB flash & GSM/HSPA/LTE\tabularnewline
(mobile client) & Octa SoC & $4$\,GB RAM & IEEE 802.11n, 2.4/5\,GHz\tabularnewline
\hline 
\end{tabular}}
\end{table}

Meanwhile, the core infrastructure host is a Windows Server virtual
machine (VM). This VM runs on as ESXi 5.5 hypervisor installed bare-metal
onto an HP Proliant DL380 G5 rack-mount server. Key software running
on this VM are MongoDB (v4.2.3), Internet Information Services (v10.0),
Eclipse Mosquitto (v1.6.11), and MQTT2HTTP (v0.1). The UE is a Samsung
Galaxy S7 G930W8 running the latest supported build of Android 8.0
Oreo with TouchWiz GUI.

Continuing to $T_{{\rm DL}}$, a GET method is written for the core
service API which responds with a 28\,bytes JSON payload. This method
is invoked via barebones flutter application running on the client
timing request until core service response. Finally, the measurement
of $T_{{\rm REND}}$ is obtained as time taken to render a data-driven
GUI component on the mobile application. We designed the mobile app
to display the availability of parking spaces (see Fig.~\ref{Fig1. Smart_Parking})
in a specific lot of Sheridan's Davis campus that features 183 spots.
The application redraws equiangular polygons in Flutter Maps onto
a MapBox PolygonLayer on the client UE's GUI to illustrate the parking
availability and information.

Overall, the latency measurements were collected during business hours,
i.e., 9:00 to 17:00 on weekdays, over a period of two weeks. For benchmarking
purposes, the experiment was repeated twice more with similar parameters
for standalone LoRa connected through WLAN to the internet service
provider, and LTE-M through Bell's cellular network.

\begin{figure*}[tbh]
\begin{centering}
\subfloat[Unlicensed IoT: standalone LoRa \label{Fig6a. LoRa}]{\begin{centering}
\includegraphics[width=0.27\paperwidth]{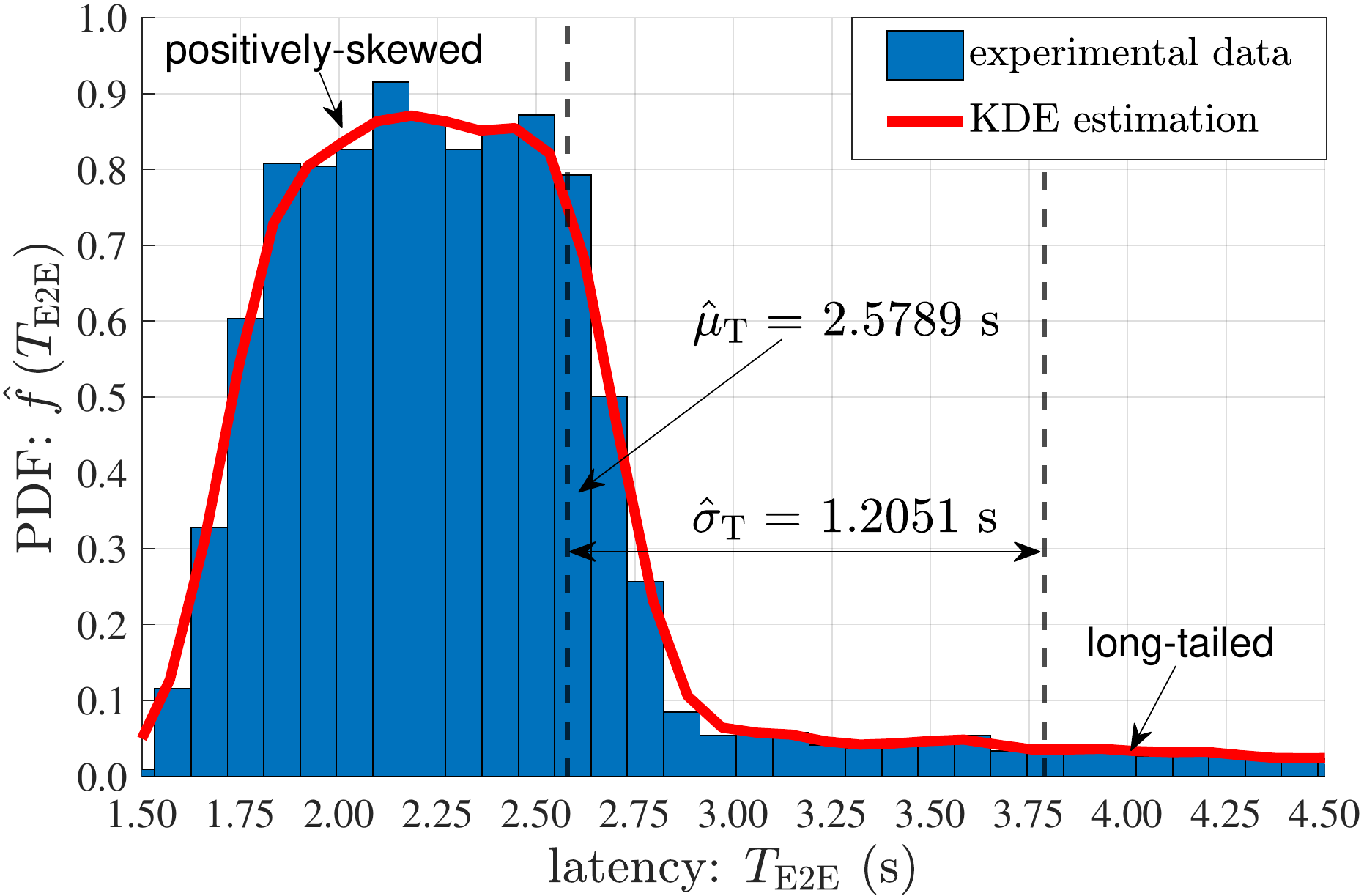}
\par\end{centering}
} \subfloat[Cellular IoT: standalone LTE-M \label{Fig6b. LTE-M}]{\begin{centering}
\includegraphics[width=0.27\paperwidth]{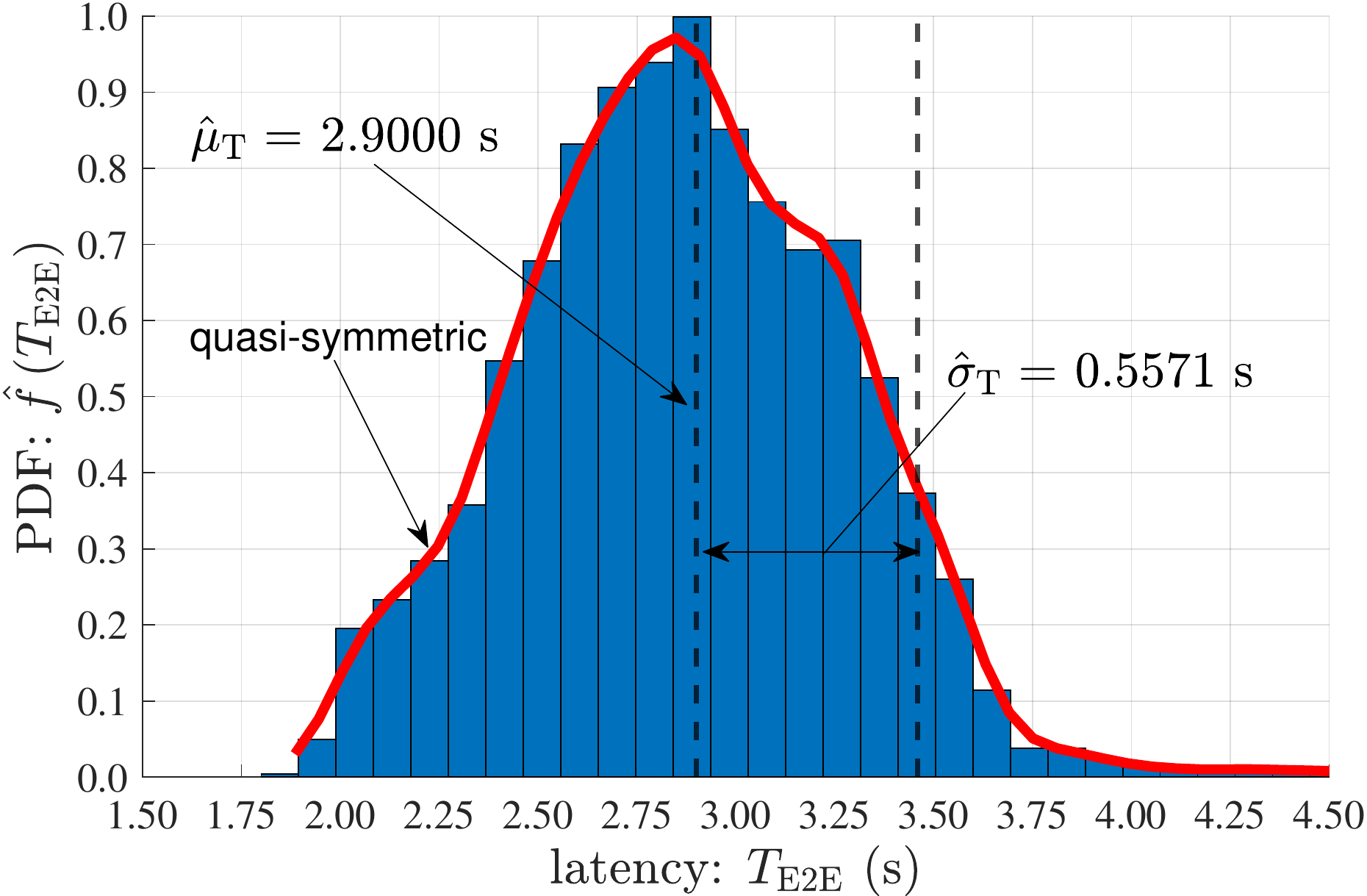}
\par\end{centering}

} \subfloat[Concatenated IoT: LoRa interfaced with LTE-M \label{Fig6c. Concatenated_LPWAN}]{\begin{centering}
\includegraphics[width=0.27\paperwidth]{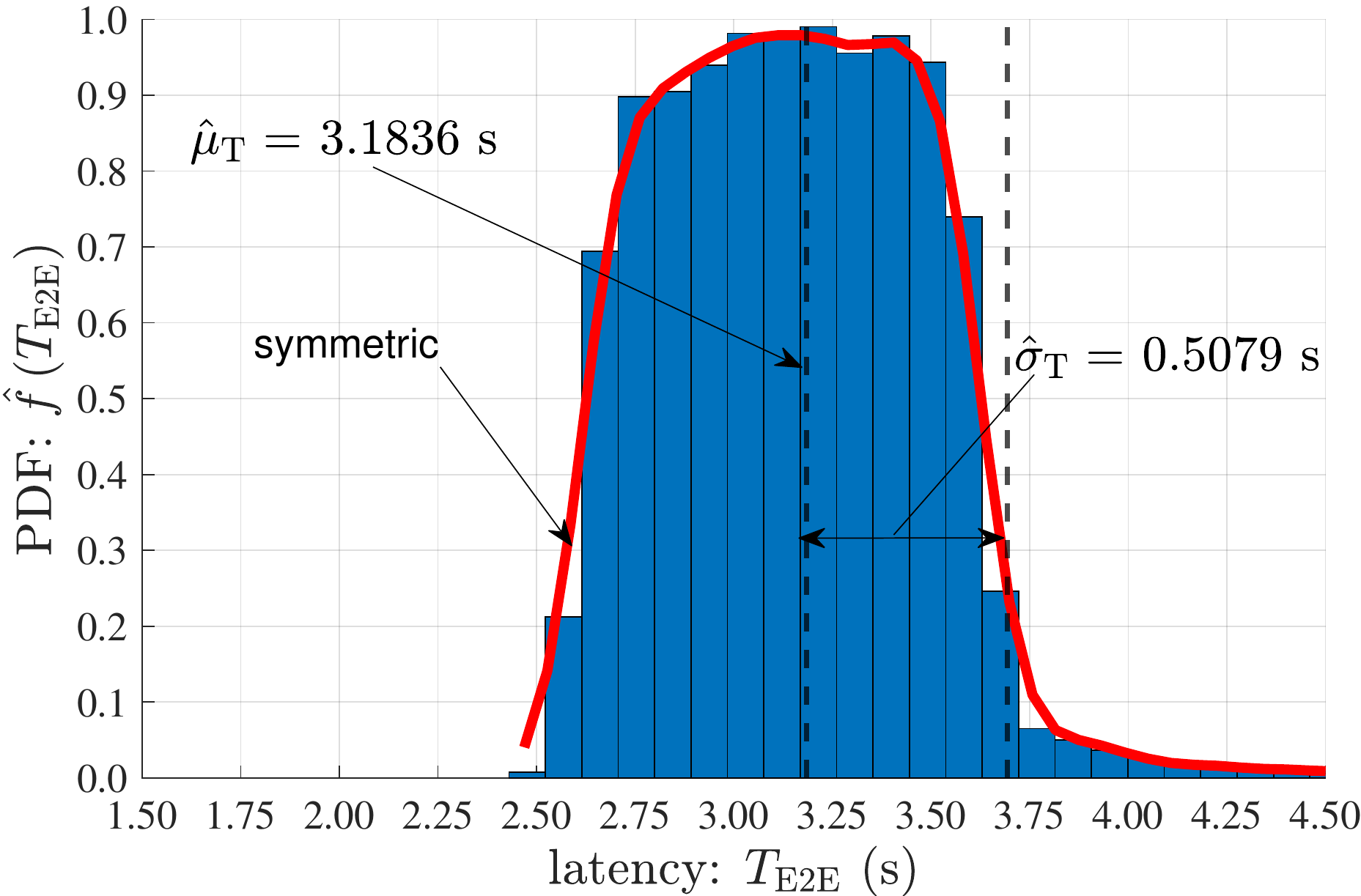}
\par\end{centering}
}
\par\end{centering}
\caption{Probability density function of E2E latency based on $n_{{\rm s}}=10,000$
experimental data points for each LPWAN network architecture scheme.
\label{Fig6. PDF_E2E_Latency_3Plots}}
\end{figure*}

\vspace{-1bp}

\vspace{-1bp}

\section{\label{Sec4 - Empirical Results}Experimental Results and Analysis
}

\vspace{-1.7bp}

In order to yield reliable statistics, in excess of $30,000$ E2E
latency measurements are experimentally collected for the standalone
and concatenated LPWAN architecture schemes. Using the dataset that
is openly available in \cite{Dataset2021}, and examining the order
of magnitude of the time measures in (\ref{eq1: Latency}), we notice
that $T_{{\rm UL}}\negthinspace\gtrapprox\negthinspace T_{{\rm DL}}\negthinspace\thickapprox\negthinspace T_{{\rm REND}}\negthinspace\gg\negthinspace T_{{\rm Q}}$.
The probability density function (PDF) of the aggregated E2E latency
$T_{{\rm E2E}}$ for unlicensed IoT (LoRa), cellular IoT (LTE-M),
and concatenated LPWAN (LoRA interfaced with LTE-M) are respectively
plotted in Fig.~\ref{Fig6. PDF_E2E_Latency_3Plots}. The $\hat{f}\left(T_{{\rm E2E}}\right)$
histogram plots are obtained with a statistically significant sample
size of $n_{{\rm S}}=10,000$ data points for each LPWAN scheme, and
$n_{{\rm B}}=150$ bins is considered for distribution accuracy (see
blue histogram bars in Fig.~\ref{Fig6. PDF_E2E_Latency_3Plots}).

In addition, the empirical distributions of $T_{{\rm E2E}}$ are approximated
analytically using the kernel density estimation (KDE),
\begin{align}
 & \hat{f}_{{\rm kde}}\left(T_{{\rm E2E}}\right)=\frac{1}{n_{{\rm S}}\thinspace h}\ \sum_{i=1}^{n_{{\rm S}}}K_{{\rm g}}\negthinspace\negthinspace\left(\frac{T_{{\rm E2E}}-T_{i}}{h}\right)\ \ \ \ \ h>0\label{eq2: KDE}
\end{align}
where a Gaussian kernel function $K_{{\rm g}}\left(t\right)=e^{-t^{2}/2}/\sqrt{2\pi}$
is used (see red curves in Fig.~\ref{Fig6. PDF_E2E_Latency_3Plots}).
The bandwidth (BW) of KDE is determined using Silverman's rule, i.e.,
$h\simeq\hat{\sigma}_{{\rm T}}^{\prime}\sqrt[5]{4/3n_{{\rm s}}}$,
where $\hat{\sigma}_{{\rm T}}^{\prime}=med\left(\left\{ \left|T_{i}-med\left(\left\{ T_{i}\right\} \right)\right|\right\} \right)/0.6745$
is the median absolute deviation (MAD), with $med\left(T\right)$
representing the median of the finite set $T=\left\{ T_{i}\right\} $,
$i=1,2,\ldots,n_{{\rm S}}$ for E2E latency measurements. MAD is
a variation metric that provides a more robust estimation for the
standard deviation (SD) through greater resiliency to outliers and
extreme values within the dataset. As evident in Fig.~\ref{Fig6. PDF_E2E_Latency_3Plots},
the KDE closely match the experimental plots for the three LPWAN architecture
schemes, where $\widetilde{\sigma}_{{\rm T}}$ is the SD for the smooth
KDE curves.

The KDE can be used to analytically characterize the randomness of
the underlying E2E experimental latencies, which is paramount in predicting
and measuring end-users' QoE. A larger sample size of latency measurements
from PoC experiments will certainly yield a more robust analytical
predictor that can be used to randomly generate reliable E2E latencies
for performance evaluation of networks.

The statistics and estimation parameters from the density plots are
shown in Table~\ref{ Tab II. Statistics of empirical data}, where
the sample mean and SD (see vertical lines in Fig.~\ref{Fig6. PDF_E2E_Latency_3Plots})
are accordingly evaluated by
\begin{align}
 & \hat{\mu}_{{\rm T}}=\frac{1}{n_{{\rm s}}}{\displaystyle \sum_{i=1}^{n_{{\rm s}}}T_{i}}\text{ ;}\thinspace\thinspace\thinspace\thinspace\thinspace\thinspace\thinspace\hat{\sigma}_{{\rm T}}=\sqrt{\frac{1}{n_{{\rm s}}-1}{\displaystyle \sum_{i=1}^{n_{{\rm s}}}\left|T_{i}-\hat{\mu}_{{\rm T}}\right|^{2}}}\text{ .}\label{eq3: Mean + SD}
\end{align}
From the results, we observe that the concatenated LPWAN architecture
has an excess latency $\Delta\hat{\mu}_{{\rm T}}$ that is on average
$284$\,ms more than the standalone cellular\,IoT with LTE-M communications.
In fact, the excess delay was expected as a consequence of interfacing
unlicensed LPWAN with cellular IoT, as this introduces more data transmission
hops on the path to the core service. Although the comparison of the
concatenated scheme to cellular IoT is more relevant, it is still
worthwhile to notice that the excess latency with LoRa over WLAN is
nearly double the average value with LTE-M.

\begin{table}[t]
\caption{Statistics and density estimation of empirical E2E latencies\label{ Tab II. Statistics of empirical data}}

\resizebox{0.50\textwidth}{!}{
\begin{tabular}{c||cc||ccc}
\hline 
\textbf{\textcolor{black}{LPWAN}} & \textbf{\textcolor{black}{mean}} & \textbf{\textcolor{black}{SD}} & \textbf{\textcolor{black}{MAD}} & \textbf{\textcolor{black}{KDE BW}} & \textbf{\textcolor{black}{SD of KDE }}\tabularnewline
\textbf{\textcolor{black}{schemes}} & \textcolor{black}{$\hat{\mu}_{{\rm T}}$ (s)} & \textcolor{black}{$\hat{\sigma}_{{\rm T}}$ (s)} & \textcolor{black}{$\hat{\sigma}_{{\rm T}}^{\prime}$ (s)} & \textcolor{black}{$h$} & \textcolor{black}{$\widetilde{\sigma}_{{\rm T}}$ (s)}\tabularnewline
\hline 
\hline 
\textcolor{black}{LoRa/LoRaWAN} & \textcolor{black}{$2.5789$} & \textcolor{black}{$1.2051$} & \textcolor{black}{$0.4321$} & \textcolor{black}{$0.0725$} & \textcolor{black}{$1.2072$}\tabularnewline
\hline 
\textcolor{black}{3GPP LTE Cat M1} & \textcolor{black}{$2.9000$} & \textcolor{black}{$0.5571$} & \textcolor{black}{$0.4221$} & \textcolor{black}{$0.0709$} & \textcolor{black}{$0.5616$}\tabularnewline
\hline 
\textcolor{black}{LoRa + LTE-M} & \textcolor{black}{$3.1836$} & \textcolor{black}{$0.5079$} & \textcolor{black}{$0.3831$} & \textcolor{black}{$0.0643$} & \textcolor{black}{$0.5119$}\tabularnewline
\hline 
\end{tabular}}\vspace{-0.3cm}
\end{table}

Moreover, it is interesting to remark that the spread of LPWAN schemes
with cellular IoT is approximately half that of unlicensed IoT with
LoRa technology. This essentially means that E2E latencies with cellular
IoT, irrespective of being standalone or concatenated, have lower
entropy. In other words, these LPWAN networks that are dependent on
mobile operators have $T_{{\rm E2E}}$ values that are more predictable.

The shape of the E2E latency PDFs in Fig.~\ref{Fig6. PDF_E2E_Latency_3Plots}
are also unique from one LPWAN network to the other. In essence, this
gives us a \textit{signature} for the latency profile of each LPWAN
architecture. In particular, unlicensed IoT exhibits a long-tailed
distribution that is positively skewed with a left-leaning curve.
On the other hand, cellular IoT and concatenated LPWAN have E2E latency
distributions that are more symmetrical, where the former is more
dispersed than the latter.

\begin{figure}[t]
\begin{centering}
\includegraphics[width=0.95\columnwidth]{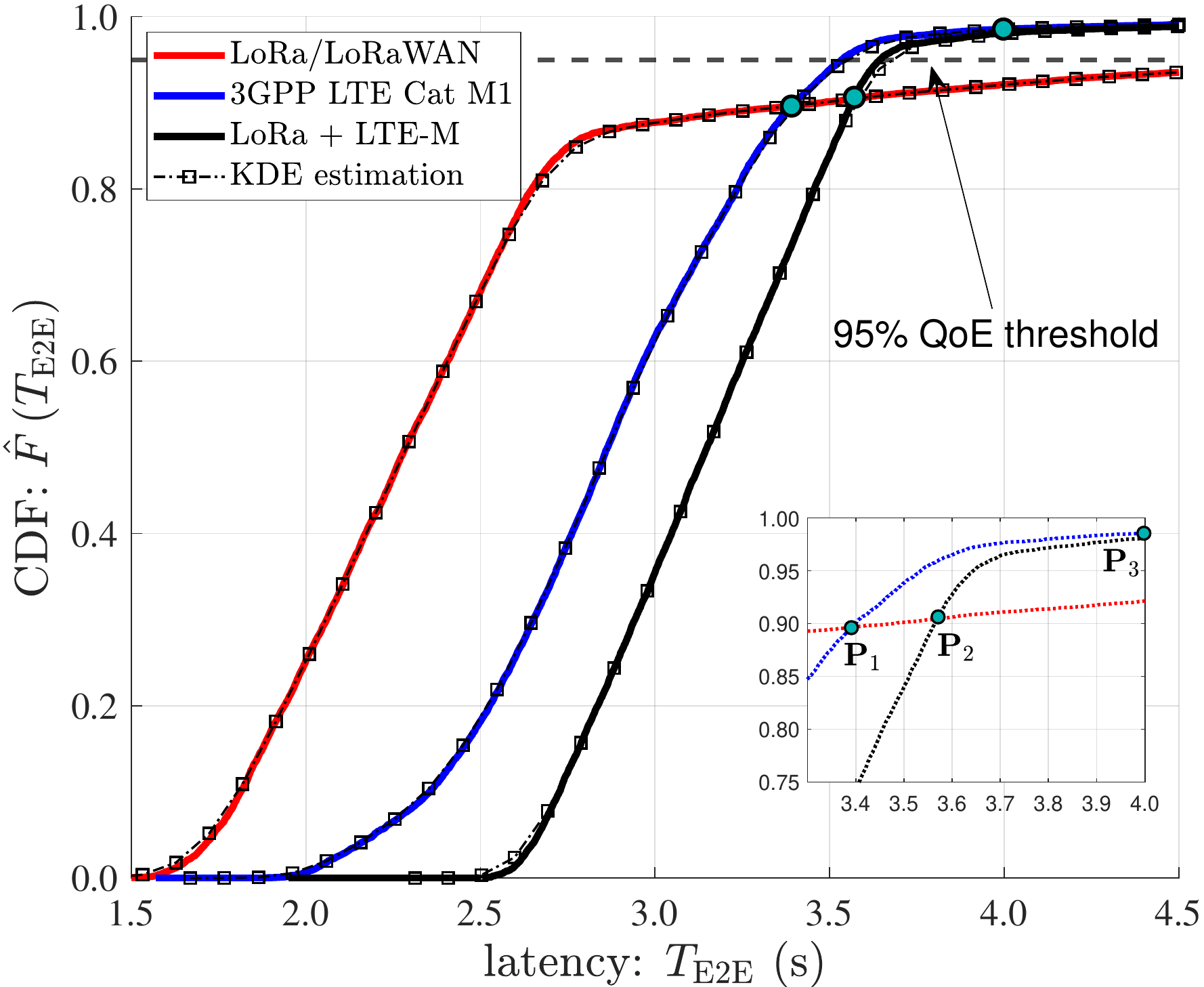}
\par\end{centering}
\vspace{-0.2cm}

\caption{Empirical CDF of E2E latency for standalone and concatenated IoT.
 \label{Fig7. CDF of E2E latency}}

\vspace{-0.3cm}
\end{figure}

To assess the feasibility of an LPWAN architecture to meet a certain
target QoE, it is insightful to also look at the cumulative distribution
function (CDF) of the particular KPI under study. As plotted in Fig.~\ref{Fig7. CDF of E2E latency}
for the three LPWAN schemes, the CDF is determined by $\hat{F}\left(T_{{\rm E2E}}\right)\negthinspace\negthinspace=\negthinspace\negthinspace\intop_{-\infty}^{T_{{\rm E2E}}}\hat{f}\left(\tau\right)\thinspace{\rm d\tau}$.
The E2E latency probability $\negthinspace\mathcal{P}_{T}\left(\cdot\right)\negthinspace$
that LPWAN sensors data is received at the application layer of an
edge user from the cloud can be used to assess the QoE performance.
This can be determined from the CDF, i.e., $\mathcal{P}_{T}\negthinspace\left(\tau_{{\rm target}}\right)\negthinspace=\negthinspace\Pr\left(T\leq\tau_{{\rm target}}\right)\negthinspace=\negthinspace\hat{F}\negthinspace\left(\tau_{{\rm target}}\right)$,
where $T\negthinspace\negthinspace\sim\negthinspace\negthinspace\hat{f}\left(T_{{\rm E2E}}\right)$
is the random variable for latency and $\tau_{{\rm target}}$ is
the stipulated E2E target delay.

Meanwhile, we notice that the CDF plots of KDE (see square marked
plots in Fig.~\ref{Fig7. CDF of E2E latency}) almost perfectly overlap
the curves from experimental measurements. From the plots, we also
identify intersection points that help us in distinguishing the different
regimes of the probability distributions of E2E latency for the LPWAN
architectures. These values are: $\mathbf{P}_{1}\negthinspace\negthinspace:\negthinspace\hat{F}\left(T_{{\rm E2E}}=3.391\right)=0.8961$;
$\mathbf{P}_{2}\negthinspace:\negthinspace\hat{F}\left(T_{{\rm E2E}}\negthinspace=\negthinspace3.571\right)\negthinspace=\negthinspace0.9063$;
and $\mathbf{P}_{3}\negthinspace:\negthinspace\hat{F}\left(T_{{\rm E2E}}\negthinspace=\negthinspace3.998\right)\negthinspace=\negthinspace0.9855$.
This means that if $\tau_{{\rm target}}\negthinspace\apprle\negthinspace3.4$\,s,
unlicensed LPWAN outperforms cellular IoT. For example, if a $3$\,s
E2E latency is accepted, LoRa surpasses with nearly $\negthinspace\mathcal{P}_{T}\negthinspace\left(\tau_{{\rm target}}\simeq3\right)\negthinspace\simeq\negthinspace88\%\negthinspace$
likelihood of meeting the latency target, while standalone LTE-M and
concatenated LPWAN achieve $63\%$ and $36\%$ respectively. On the
other hand, if for certain smart city use cases an extra second is
tolerated, then, LTE-M and concatenated LPWAN outperform standalone
LoRa by a factor of $6.4\%$. In this situation, LTE-M and concatenated
LPWAN both achieve the same latency performance of $99\%$, irrespective
of excess latency. However, the minimum QoE threshold for user satisfaction
should generally aim to satisfy a performance of $\negthinspace\mathcal{P}\left(T_{{\rm E2E}}\right)\geq0.95$
for real-life applications (see horizontal line in Fig.~\ref{Fig7. CDF of E2E latency}).

\vspace{-1bp}

\vspace{-1bp}

\vspace{-1bp}

\vspace{-1bp}

\section{\label{Sec5 - Conclusion}Conclusion }

\vspace{-1bp}

In this paper, we proposed a concatenated LPWAN architecture over
a private core infrastructure that interfaces LoRa with cellular IoT
in order to mitigate poor QoS and costly service plans. Through precise
QoE assessment, E2E latency measurements were experimentally recorded
for the concatenated architecture, and the data was compared to standalone
licensed and unlicensed LPWANs. The statistics from the density functions
reveal that the proposed concatenated LPWAN added, on average, an
excess latency of $23.5\%$ and $9.8\%$ when compared to standalone
LoRa and LTE-M. Nevertheless, such milliseconds of E2E excess latency
is humanly insignificant for use cases where perception is needed.
More importantly, we also found that concatenated LPWAN outperforms
unlicensed IoT by roughly $1.5$\,s at the typical QoE threshold
of $95\%$ for users' satisfaction. Overall, this experimental study
suggests that a concatenated LPWAN is technically feasible and offers
an affordable alternative for real-world smart city deployment.

\vspace{-1bp}

\section*{Acknowledgments}

\vspace{-1bp}

This research work is supported, in part, by Sheridan\textquoteright s
Research and Creative Activities Growth Grant, and by Bell Canada\textquoteright s
IoT Solutions for eMTC (3GPP LTE Cat M1) network access and services.
The authors are also grateful to ENCQOR-5G program for testbed access.
They also would like to acknowledge Keysight Technologies for research
discussions regarding implementation and measurements.

\vspace{-1bp}

\bibliographystyle{IEEEtran}
\bibliography{IEEEabrv,IEEEexample,REF_VTC21}

\end{document}